\documentstyle[aps,prl,preprint,floats,epsf]{revtex}
\newcommand{\lnu}{\ell\bar{\nu}}
\newcommand{\dlnu}{D\ell\bar{\nu}}
\newcommand{\dplnu}{D^+\ell^-\bar{\nu}}
\newcommand{\dzlnu}{D^0\ell^-\bar{\nu}}
\newcommand{\dslnu}{D^*\ell\bar{\nu}}
\newcommand{\ddlnu}{D^{**}\ell\bar{\nu}}
\newcommand{\Vcb}{|V_{cb}|}
\newcommand{\bo}{B^0}
\newcommand{\bob}{\bar{B}^0}
\newcommand{\bp}{B^+}
\newcommand{\bm}{B^-}
\newcommand{\barb}{\bar{B}}
\newcommand{\bb}{B\bar{B}}
\newcommand{\Ufs}{\Upsilon(4S)}
\newcommand{\kpi}{K^-\pi^+}
\newcommand{\kpipi}{K^-\pi^+\pi^+}

\newcommand{\de}{\Delta E}

\newcommand{\mcand}{M_{\text{cand}}}
\newcommand{\fw}{{\cal F}_D(w)}
\newcommand{\fsw}{{\cal F}_{D^*}(w)}
\newcommand{\fone}{{\cal F}_D(1)}
\newcommand{\rh}{\hat{\rho}^2_D}

\draft

\begin{document}
%
%
%%%%%%%%%%%%%%%%%%%%%%%%%%%%%%%%%%%%%%%%%%%%%%%%%%
%                                                %
%    Title page info                             %
%                                                %
%%%%%%%%%%%%%%%%%%%%%%%%%%%%%%%%%%%%%%%%%%%%%%%%%%

\preprint{\tighten\vbox{\hbox{CLNS 97-1486 \hfill}
                        \hbox{CLEO 97-12   \hfill}}}

\title{Measurement of the $\barb\to\dlnu$ Partial Width and Form
Factor Parameters}

\date{\today}

\maketitle

\begin{abstract}
\tighten
We have studied the decay $\barb\to\dlnu$, where $\ell=e\text{ or }
\mu$.  From a fit to the differential decay rate $d\Gamma/dw$ we
measure the rate normalization $\fone\Vcb$ and form factor slope
$\rh$, and, using measured values of $\tau_B$, find
$\Gamma(\barb\to\dlnu) = (12.0 \pm 0.9 \pm 2.1) \ {\rm ns^{-1}}$.  The
resulting branching fractions are ${\cal B}(\bob\to\dplnu)=(1.87 \pm
0.15 \pm 0.32)\%$ and ${\cal B}(\bm\to\dzlnu)=(1.94 \pm 0.15 \pm
0.34)\%$.  The form factor parameters are in agreement with those
measured in $\barb\to\dslnu$ decays, as predicted by heavy quark
effective theory.
\end{abstract}


\begin{center}
\tighten
M.~Athanas,$^{1}$ P.~Avery,$^{1}$ C.~D.~Jones,$^{1}$
M.~Lohner,$^{1}$ C.~Prescott,$^{1}$ J.~Yelton,$^{1}$
J.~Zheng,$^{1}$
G.~Brandenburg,$^{2}$ R.~A.~Briere,$^{2}$ A.~Ershov,$^{2}$
% ...
\end{center}
 
{\em \small
\begin{center}
\tighten
$^{1}${University of Florida, Gainesville, Florida 32611}\\
$^{2}${Harvard University, Cambridge, Massachusetts 02138}\\
% ...
$^{12}${University of Oklahoma, Norman, Oklahoma 73019}\\
$^{13}${Purdue University, West Lafayette, Indiana 47907}\\
$^{14}${University of Rochester, Rochester, New York 14627}\\
$^{15}${Stanford Linear Accelerator Center, Stanford University, Stanford,
California 94309}\\
% ...
$^{20}${Wayne State University, Detroit, Michigan 48202}\\
% ...
\end{center}}
\clearpage 
\narrowtext
%
%
%%%%%%%%%%%%%%%%%%%%%%%%%%%%%%%%%%%%%%%%%%%%%%%%%%
%                                                %
%    BEGINNING OF TEXT                           %
%                                                %
%%%%%%%%%%%%%%%%%%%%%%%%%%%%%%%%%%%%%%%%%%%%%%%%%%

\tighten

Exclusive semileptonic $B$ meson decays provide information about both
the weak and strong interactions of quarks.  The rate for these decays
is proportional to the square of the CKM matrix element
$\Vcb$~\cite{bb:CKM}, while the dynamics of these decays, as expressed
in the decay form factors, provide information about the QCD potential
which binds quarks together as hadrons~\cite{bb:ISGW,bb:formfac}.
Heavy quark effective theory (HQET)~\cite{bb:HQET} has made it
possible to extract values of $\Vcb$, with relatively little model
dependence, through a measurement of the decay rate at the point of
zero recoil of the daughter meson.  The decay $\barb\to\dslnu$ has
provided measurements of the decay rate at the point of zero recoil
and the form factor slope, giving very accurate values of
$\Vcb$~\cite{bb:dslnures} and information about the shape
of the Isgur-Wise function that describes the dynamics of
heavy-to-heavy meson transitions.

The pseudoscalar-to-pseudoscalar decay $\barb\to\dlnu$ can provide the
same information, although it will be less precise because of the
smaller overall decay rate for this mode than for $\barb\to\dslnu$,
the smaller rate near the point of zero recoil, and ${\cal O}(1/M_Q)$
corrections to the decay rate at that point that are not present in
$\barb\to\dslnu$.  However, this mode should yield a value of $\Vcb$ that
is consistent with other measurements, and HQET
predicts~\cite{bb:dlnuthr} that the form factor parameters should be
very nearly the same as in the $\barb\to\dslnu$ decay.  In this
Letter, we present a measurement of the differential decay rate
$d\Gamma/dw$ in the decay $\barb\to\dlnu$, from which we extract the
decay rate normalization at the point of zero recoil and the form
factor slope.  The variable $w=(M_B^2 + M_D^2 - q^2)/(2M_B M_D)$,
where $q^2$ is the invariant mass squared of the lepton-neutrino
system, is the kinematic variable of HQET, and is equal to the
relativistic $\gamma$ factor for the $D$ meson in the $B$ meson rest
frame.  From these results and the measured $B$ lifetime, we
obtain the partial width for the decay and convert it to branching
fractions.

This study is based on an $\Ufs$ data sample of 3.16 ${\rm fb^{-1}}$
(3.34 $\times 10^6\ \bb$ pairs) accumulated by the CLEO experiment at
the Cornell Electron Storage Ring (CESR).  The CLEO
detector~\cite{bb:CLEO-nim} contains three concentric wire chambers
that detect charged particles and a CsI(Tl) electromagnetic
calorimeter that detects photons, all within a 1.5 T superconducting
solenoid.  

The undetected neutrino complicates analysis of semileptonic decays.
Using the hermeticity of the CLEO detector, we reconstruct the
neutrino by inferring its four-momentum from the missing energy
($E_{\text{miss}}\equiv 2E_{\text{beam}}-\sum E_i$) and missing
momentum ($\vec{p}_{\text{miss}}\equiv -\sum\vec{p}_i$) in each event,
where $E_i$ and $\vec{p}_i$ are the energy and momentum of each detected
particle $i$ in the event, as was done in the CLEO measurement of
exclusive $b\to u\ell\bar{\nu}$ decay rates~\cite{bb:CLEO-pilnu}.  In
the process $e^+e^-\to\Upsilon(4S)\to B\bar{B}$, the total energy of
the beams is imparted to the $B\bar{B}$ system.  At CESR, that system
is at rest, so the neutrino combined with the signal lepton and $D$
meson should satisfy the energy constraint $\Delta E \equiv
(E_{\bar{\nu}}+E_\ell+E_D)-E_{\text{beam}} = 0$ and the momentum
constraint $\mcand \equiv [E_{\text{beam}}^2 -
|\vec{p}_{\bar{\nu}}+\vec{p}_\ell+\vec{p}_D|^2]^\frac12 = M_B$.  We
select candidates with $5.2650 \leq \mcand < 5.2875$~GeV and $-100 \leq
\de < 500$~MeV; the requirement on $\de$ is asymmetric about zero to
reject feeddown from $\barb\to\dslnu$ decays, which have $\de$ values
of about --150~MeV when reconstructed as $\barb\to\dlnu$ decays.

To suppress events in which $\vec{p}_{\text{miss}}$ misrepresents
$\vec{p}_{\bar{\nu}}$, we reject those events with multiple leptons or
a total charge more than one unit from zero because they indicate
other missing particles.  We further require that $M^2_{\text{miss}}
\equiv E_{\text{miss}}^2 - |\vec{p}_{\text{miss}}|^2$ for each event
be consistent with zero.  Surviving signal events show a resolution in
$|\vec{p}_{\text{miss}}|$ of 110~MeV/$c$.  Because the resolution on
$E_{\text{miss}}$ is about 2.1 times larger, we take
$(E_{\bar{\nu}},\vec{p}_{\bar{\nu}})=
(|\vec{p}_{\text{miss}}|,\vec{p}_{\text{miss}})$.

Information from calorimeter and tracking measurements including
specific ionization is combined to identify electrons with
$p>600\text{ MeV}/c$ over 90\% of the solid angle.  Particles are
considered muons if they register hits in counters deeper than 5
interaction lengths over the polar angle range $|\cos\theta|<0.85$.
Candidate leptons must have $0.8 \leq p_\ell <2.4 \text{ GeV}/c$, where the
lepton identification efficiency averages more than 90\%; the probability
that a hadron is misidentified as an electron (muon), a ``fake
lepton'', is about 0.1\% (1\%).

The leptons and neutrinos are then combined with $D$ mesons, which are
identified in the decay modes $D^0\to \kpi$ and $D^+\to\kpipi$.  (The
charge conjugate mode is always implied.)  Hadron mass assignments are
made by requiring that the kaon and lepton have the same charge.  To
reduce large backgrounds to $D^+\to\kpipi$ decays from random track
combinations, we require that the $D^+$ daughter pions (kaon) have
measured specific ionization values consistent with the assumed
particle hypothesis within 3.5 (3) standard deviations.  $D^0$
candidates are required to have invariant mass satisfying $1.850 \leq
M_{K\pi} < 1.880$~GeV, and $D^+$ candidates to satisfy $1.855 \leq
M_{K\pi\pi} < 1.885$~GeV.  This $\pm$15~MeV range is about twice the
experimental resolution on the invariant mass.  Since the $D$ must
come from a $B$ decay, we require $p_D<2.6$~GeV/c.  We also require
that the angle between the directions of the $D$ candidate and the
lepton-neutrino system have a physical value;the angle is calculated
from the $D$ meson and lepton momenta and the beam energy only,
without examining the missing momentum.

\begin{figure}
\centering
\leavevmode
\epsffile{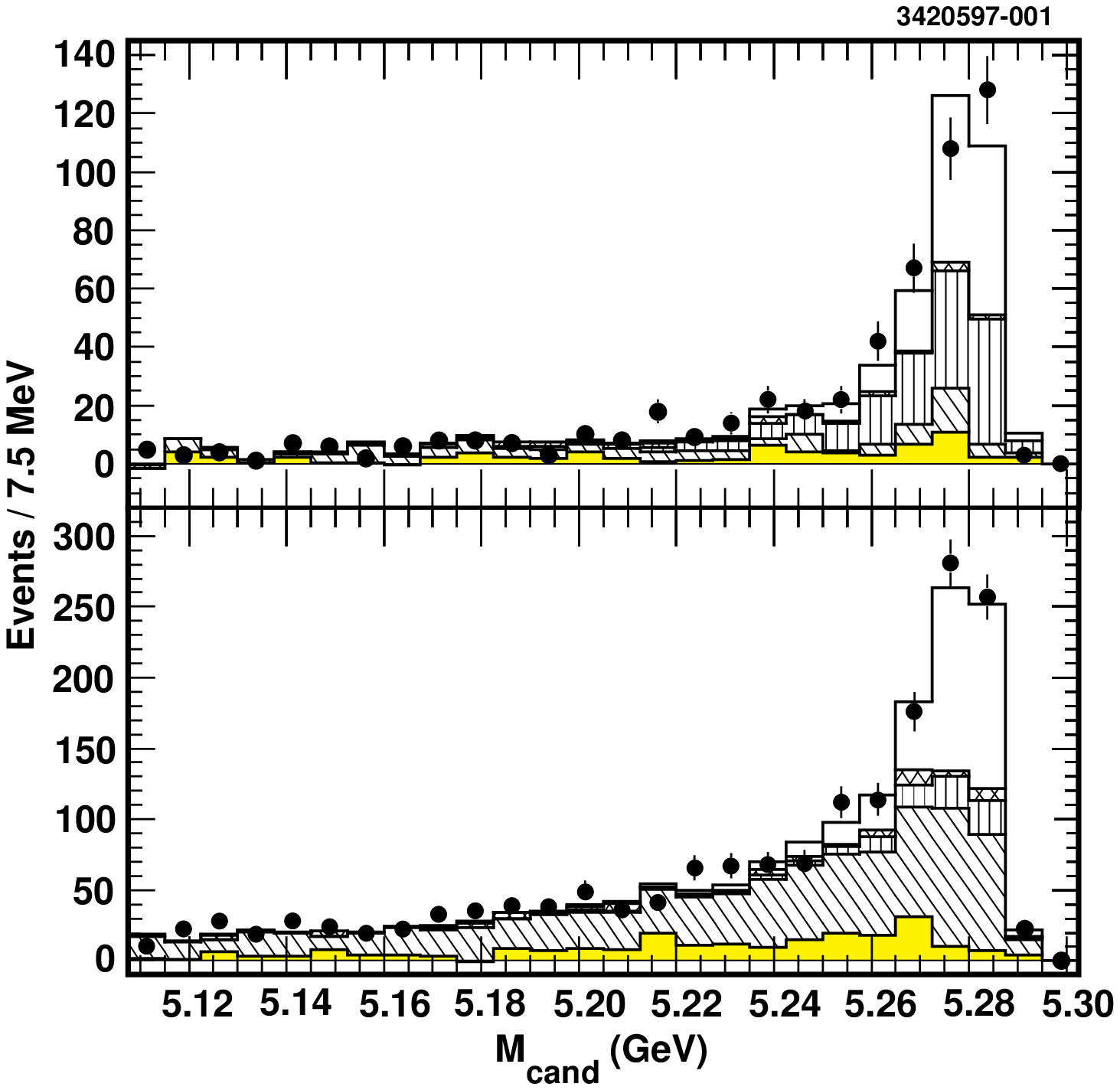}
\caption{$\mcand$ distributions for the $\dzlnu$ (top) and $\dplnu$
(bottom) modes.  The points are the data, the shaded component is the
continuum and fake lepton background, the diagonal hatch is the
combinatoric background, the vertical hatch is the $\barb\to\dslnu$
feeddown, and the crosshatch is the $\barb\to\ddlnu$ and other
backgrounds.  The unshaded area is the prediction of a Monte Carlo
simulation of the signal, normalized to the measured decay rate.}
\label{fig:bmass}
\end{figure}

Backgrounds arise from $e^+e^-\to q\bar{q}/\tau^+\tau^-$ (continuum),
fake leptons, random track combinations that form $D$ candidates,
feeddown from other $\barb\to DX\ell\bar{\nu}$ decays, and random
combinations of $D$ candidates and leptons from different parent $B$
mesons.  The continuum backgrounds are reduced by requiring that the
ratio of Fox-Wolfram moments $H_2/H_0$~\cite{bb:FW} be less than
0.4. Backgrounds from $\barb\to D^*X\ell\bar{\nu}$ decays are reduced
by eliminating events that include $D\pi$ or $D\gamma$ pairs that are
consistent with $D^*$ decay.
We find 303 $\dzlnu$ and 714 $\dplnu$ candidates that satisfy all of these
requirements.  The average reconstruction efficiency, as determined by a
Monte Carlo simulation of the CLEO detector, is 2.82\% for
$\bm\to\dzlnu$ events and 2.57\% for $\bob\to\dplnu$ events.
Figure~\ref{fig:bmass} shows the $\mcand$ distribution observed in the
data.

We estimate the continuum background using data collected 60~MeV below
the $\Upsilon(4S)$ energy and the fake lepton background by applying
measured fake rates to nonleptonic data.  We estimate backgrounds from
random track combinations that form $D$ candidates using events in
sideband regions on either side of the $\kpi(\pi^+)$ invariant mass
signal region under the assumption that the magnitude of this
background is linear in $\kpi(\pi^+)$ invariant mass.  This is the
largest background in the $\dplnu$ sample.

Feeddown backgrounds are modeled through Monte Carlo simulations,
using an event generator that accounts for all angular correlations
among the decay products, and a full simulation of the CLEO detector.
The magnitude of the $\barb\to\dslnu$ background is normalized to the
measured rate for this decay~\cite{bb:PDG}.  This is the largest
background in the $\dzlnu$ sample, as the kinematics of the signal and
background decays are so similar, and because $D^*$ mesons are more
likely to decay to a $D^0$ than to a $D^+$.
$\barb\to\ddlnu$ processes, where $D^{**}$ represents a variety of
charm mesons with radial and angular excitations and nonresonant
$D^{(*)}\pi$ states, are modeled with the ISGW2~\cite{bb:ISGW} and
Goity and Roberts~\cite{bb:GR} models.  We normalize these backgrounds
to a set of rates~\cite{bb:slbr} that are consistent with existing
measurements~\cite{bb:ddlnu} and the total $B$ semileptonic
decay rate and lepton momentum spectrum~\cite{bb:Xlnu}.

Remaining backgrounds, such as those from random combinations of $D$
mesons and leptons and those from misreconstructed $D$ mesons not
accounted for in the combinatoric background estimate, are modeled by
a Monte Carlo simulation that reproduces the general features of
$B$-meson decay, including the inclusive-lepton and $D$ momentum
distributions.  The magnitudes of the various backgrounds are
summarized in Table~\ref{tab:events}.  Figure~\ref{fig:bmass} shows
the $\mcand$ distributions of the backgrounds.  After accounting for
these backgrounds in our sample, the lepton momentum and decay angle
distributions for selected events are consistent with $\barb\to\dlnu$
decays.

\begin{table}[t]
\widetext
\caption{Event yields and background estimates.  Errors are
statistical only.}
\label{tab:events}
\begin{tabular}{ccc}
& $\dzlnu$ & $\dplnu$ \\ \hline
Total Yield & 303.0 $\pm$ 17.4 & 714.0 $\pm$ 26.7\\\hline
Continuum & 17.3 $\pm$ 5.8 & 46.2 $\pm$ 9.4\\
Combinatoric & 26.1 $\pm$ 4.6 & 256.5 $\pm$ 10.8 \\
Fake lepton & 2.9 & 2.3 \\
$\barb\to\dslnu$ & 107.2 & 62.6 \\
$\barb\to\ddlnu$ & 5.3 & 9.9 \\
Other & --0.1 & 13.6 \\ \hline
Corrected Yield & 144.3 $\pm$ 18.9 & 322.8 $\pm$ 30.3 \\
\end{tabular}
\narrowtext
\end{table}

\begin{figure}
\centering
\leavevmode
\epsffile{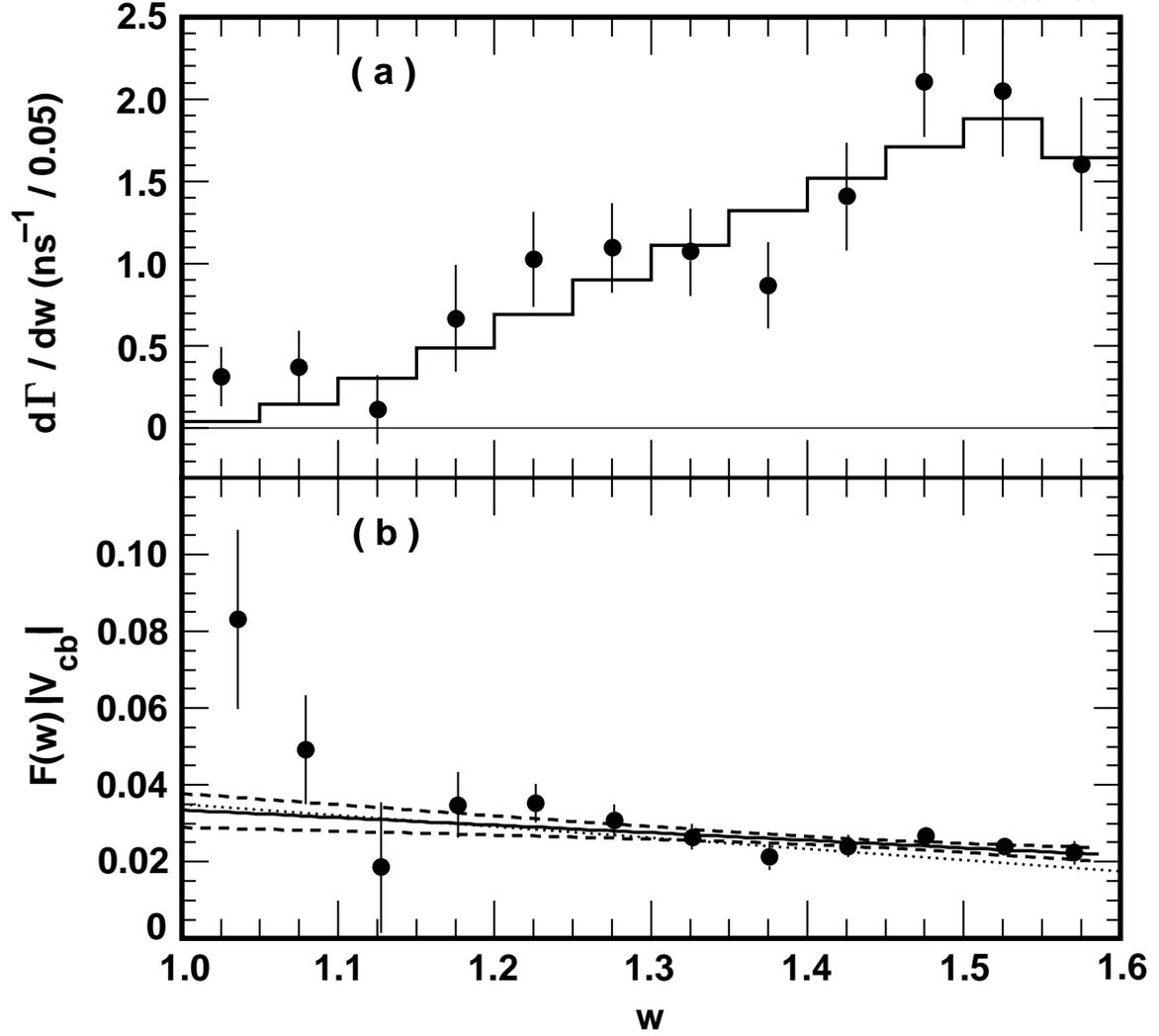}
\caption{(a) $d\Gamma/dw$ distribution from the data (points), and the
result of the $c=0$ fit to the distribution (histogram).  (b) Measured
values of $\fw\Vcb$ (points), the result of the fit (solid line) along
with its statistical errors (dashed lines), and the function
$\fsw\Vcb$ obtained from an analysis of $\barb\to\dslnu$ decays
(dotted line).}
\label{fig:wplot}
\end{figure}

To extract the form factor parameters and partial width, we construct the
distribution
\begin{displaymath}
\frac{d\Gamma}{dw} = \frac{1}{4N_{\Ufs}}
\left[\frac{N_0(w)}{\tau_{\bm}{\cal B}_{D^0}\epsilon_0(w)} +
\frac{N_+(w)}{\tau_{\bob}{\cal B}_{D^+}\epsilon_+(w)} \right],
\end{displaymath}
where $N_{\Ufs}$ is the number of $\Ufs$ events, $\epsilon(w)$ is the
reconstruction efficiency as a function of $w$, $\tau_B$ is the $B$
lifetime, and ${\cal B}$ is the appropriate $D\to\kpi(\pi^+)$
branching fraction~\cite{bb:PDG}.  By combining the two decay modes in
this fashion and assuming that the partial widths of the two modes are
equal, the results are independent of the division of $\Ufs$ decays
between $\bp\bm$ and $\bo\bob$.  This distribution is shown in
Figure~\ref{fig:wplot}a.

The $d\Gamma/dw$ distribution for $\barb\to\dlnu$ decays is~\cite{bb:dlnuthr}
\begin{displaymath}
\frac{d\Gamma}{dw} = 
\frac{G^2_F|V_{cb}|^2}{48\pi^3} (m_B+m_D)^2 (m_D\sqrt{w^2-1})^3 \fw^2.
\end{displaymath}
The function $\fw$ is the decay form factor, which can be parameterized by
the expression
\begin{displaymath}
\fw = \fone(1-\rh(w-1)+\hat{c}_D(w-1)^2).
\end{displaymath}
In the limit of infinitely heavy quarks, $\fw$ becomes the Isgur-Wise
function $\xi(w)$, and HQET predicts that $\fone=1$; for finite mass
quarks, $\fone$ can be estimated in the framework of
HQET~\cite{bb:HQET}.  The values of $\rh$ and $\hat{c}_D$ are unknown;
many previous form factor measurements have set $\hat{c}_D=0$.  We
follow this convention, but explore our sensitivity to this assumption
using various models that provide relations between $\rh$ and
$\hat{c}_D$.  We perform a $\chi^2$ fit of our $d\Gamma/dw$ distribution
to the convolution of the expected form with a function that accounts
for detector resolution in $w$ ($\pm$0.015).  We allow two free
parameters, the rate normalization $\fone\Vcb$ and the form factor
slope $\rh$.  By integrating the fit over the entire range of $w$, we
obtain the partial width for the decay.  In the fit, shown in
Figure~\ref{fig:wplot}a, $\chi^2=11.5$ for 10 degrees of freedom, and
the correlation between the two parameters is 0.95.

Systematic errors, summarized in Table~\ref{tab:systematics}, are
dominated by uncertainty in the decay model of the non-signal $B$ and
inaccuracies in detector simulation.  These effects are investigated
by varying the $K^0_L$ fraction, charm semileptonic decay rate,
charged-particle and photon-finding efficiencies, false
charged-particle and photon simulation, charged-particle momentum
resolution, and photon-energy resolution~\cite{bb:thesis}.
Uncertainties in the feeddown background normalizations and the
$\barb\to\dslnu$ decay form factors~\cite{bb:dslnuff} have their most
significant effect on the form-factor slope.  Table~\ref{tab:models}
gives results for various models of $\fw$.  The partial width is not
sensitive to the choice of form-factor parameterization, but the
values of $\fone\Vcb$, $\rh$, and $\hat{c}_D$ are sensitive to this
choice.  We use these results to determine our model uncertainties.

\begin{table}
\widetext
\caption{Contributions to the systematic error (\%) in the fit
parameters and partial width.  Simulation of the detector and the
second $B$ meson contribute to $\nu$ simulation.}
\label{tab:systematics}
\begin{tabular}{cccc}
Source & $\Vcb$ & $\rh$ & $\Gamma$\\ \hline
$\nu$ simulation & 11.4 & 12.5 & 14.8 \\ 
Background normalization & 7.6 & 15.4 & 5.4 \\ 
$\barb\to\dslnu$ form factors & 1.8 & 7.1 & 2.0 \\ 
$\tau_B$ & 2.0 & 2.5 & 3.8 \\ 
Lepton ID & 1.0 & - & 2.0 \\ 
$K/\pi$ ID & 0.1 & 0.7 & 0.8 \\
Luminosity & 0.9 & - & 1.8 \\ 
$D$ branching fractions & 2.8 & - & 5.6 \\ \hline
Total & 14.3 & 21.2 & 17.5 \\ 
\end{tabular}
\narrowtext
\end{table}

\begin{table}
\widetext
\caption{Results for various parameterizations of $\fw$.  The
first errors are statistical and the second are systematic.
$\rh$ and $\hat{c}_D$ are entirely correlated in the
Caprini-Neubert~\protect\cite{bb:fone} and Boyd~\protect\cite{bb:Boyd}
models, and fixed in the ISGW2 model~\protect\cite{bb:ISGW}.}
\label{tab:models}
\begin{tabular}{lccccc}
Model & $\fone\Vcb/10^{-2}$ & $\rh$ & $\hat{c}_D$ &
$\Gamma\ {\rm (ns^{-1})}$ & $\chi^2$/dof \\\hline
$\hat{c}_D=0$ & 3.37 $\pm$ 0.44 $\pm$ 0.48 & 0.59 $\pm$ 0.22 $\pm$ 0.12 & 
0 & 12.0 $\pm$ 0.9 $\pm$ 2.1 & 11.5/10 \\
$\hat{c}_D$ free & 4.57 $\pm$ 1.10 $\pm$ 0.92 & 1.84 $\pm$ 0.81 $\pm$ 0.53 &
1.75 $\pm$ 1.15 $\pm$ 0.56 & 12.3 $\pm$ 1.0 $\pm$ 2.2 & 10.3/9 \\
C-N & 3.90 $\pm$ 0.65 $\pm$ 0.68 & 
1.18 $\pm$ 0.37 $\pm$ 0.23 &
0.78 $\pm$ 0.27 $\pm$ 0.17 & 12.1 $\pm$ 1.0 $\pm$ 2.2 & 10.8/10 \\
Boyd & 3.71 $\pm$ 0.60 $\pm$ 0.61 &
1.05 $\pm$ 0.38 $\pm$ 0.22 &
0.94 $\pm$ 0.43 $\pm$ 0.25 & 11.9 $\pm$ 0.9 $\pm$ 2.2 & 11.1/10 \\
ISGW2 & 3.25 $\pm$ 0.13 $\pm$ 0.27 & 0.64 &
0.61 & 12.1 $\pm$ 1.0 $\pm$ 2.0 & 11.6/11 \\
\end{tabular}
\narrowtext
\end{table}

Our final results are
\begin{eqnarray*}
\fone\Vcb & = & (3.37 \pm 0.44 \pm 0.48^{+0.53}_{-0.12}) \times 10^{-2} \\
\rh & = & 0.59 \pm 0.22 \pm 0.12^{+0.59}_{-0} \\
\Gamma(\barb\to\dlnu) & = & (12.0 \pm 0.9 \pm 2.1)\ {\rm ns^{-1}},
\end{eqnarray*}
where the first two errors are statistical and systematic, and
the third arises from the form-factor model variations.  This partial
width leads to branching fractions of
\begin{eqnarray*}
{\cal B}(\bob\to\dplnu) & = & (1.87 \pm 0.15 \pm 0.32)\% \\
{\cal B}(\bm\to\dzlnu) & = & (1.94 \pm 0.15 \pm 0.34)\%,
\end{eqnarray*}
where the errors are completely correlated between the two branching
fractions.  We obtain consistent results when the two decay modes are
treated separately.  Taking $\fone = 0.98 \pm 0.07$~\cite{bb:fone}, we
find $\Vcb = (3.44 \pm 0.45 \pm 0.49^{+0.54}_{-0.12} \pm 0.25) \times
10^{-2}$, where the last error arises from the uncertainty in $\fone$.
This value of $\Vcb$ is consistent with those obtained by other
means~\cite{bb:dslnures,bb:othervcb}.
Figure~\ref{fig:wplot}b shows the measured values of $\fw\Vcb$ and the
result of the fit, along with the function $\fsw\Vcb$ as measured in
$\barb\to\dslnu$ decays by CLEO~\cite{bb:dslnures}.  In comparing
these two measurements, we find $\fone/{\cal F}_{D^*}(1) = 0.96 \pm 0.20$
and $\rh - \hat{\rho}_{D^*}^2 = -0.25 \pm 0.29$.  As predicted by HQET,
the two form factors have similar normalizations and slopes.

In summary, we have measured the rate normalization and form-factor
slope in $\barb\to\dlnu$ decays.  The resulting partial width leads to
branching fractions for the charged and neutral $B$ decay modes.  The
measured form factor parameters are consistent with those measured in
$\barb\to\dslnu$ decays, as predicted by HQET, and the value of $\Vcb$
is consistent with other measurements.

We gratefully acknowledge the effort of the CESR staff in providing us
with excellent luminosity and running conditions.  This work was
supported by the National Science Foundation, the U.S. Department of
Energy, the Heisenberg Foundation, the Alexander von Humboldt
Stiftung, Research Corporation, the Natural Sciences and Engineering
Research Council of Canada, and the A.P. Sloan Foundation.

\nopagebreak


\begin{thebibliography}{99}
\bibitem{bb:slbr} We assume the following set of decay rates:
${\cal B}(\barb\to D_1\lnu)=0.60\%$,
${\cal B}(\barb\to D_2^*\lnu)=0.79\%$,
${\cal B}(\barb\to D_1^*\lnu)=0.26\%$,
${\cal B}(\barb\to D_0^*\lnu)=0.26\%$,
${\cal B}(\barb\to D'\lnu)=0.26\%$,
${\cal B}(\barb\to {D^{*}}'\lnu)=0.53\%$,
${\cal B}(\barb\to (D\pi)\lnu)=0.76\%$, and
${\cal B}(\barb\to (D^*\pi)\lnu)=0.24\%$, where $(D\pi)$ and $(D^*\pi)$
indicates non-resonant production.
\end{thebibliography}
\end{document}